\documentstyle[preprint,aps,pra,eqsecnum]{revtex}
\begin{document}
\bibliographystyle{unsrt}
\title {A new Hamiltonian  for a massive relativistic particle with spin one in a generalized Heisenberg/Schr\"odinger picture}  
\date{\today}
\author{
Rudolf A. Frick\thanks{Email: rf@thp.uni-koeln.de}\\{\it Institute of Theoretical Physics, University of Cologne, D-50923 Cologne, Germany} \\}

\maketitle
 
\begin{abstract}
 We consider a particular 4-dimensional generalization of the transition from the 
Heisenberg        to the Schr\"odinger picture. The space-time independent  expansion  with 
respect to the unitary irreducible representations of the Lorentz group is  applied, as Fourier transformation
 in the Heisenberg picture, to  the states of a massive relativistic particle. 
A new Hamiltonian  operator has been found for such a particle with spin one.
\end{abstract}

\bigskip

\newpage

\section{Introduction}

    In this paper we present a new mathematical formalism for describing a massive relativistic particle with spin one. In this formalism, we use a  four-dimensional transition from the  Heisenberg to the Schr\"odinger picture. In  quantum mechanics, the transition from the Heisenberg   to the Schr\"odinger picture is 
carried out by the unitary transformation $S(t)=\exp(-itH)$, where ${H}$ is the Hamiltonian operator of the particle; (we choose here a system of units such that ${\hbar}=1,{\,}c=1).$ 
 The state of a particle in the Heisenberg picture  and the particle operators in the 
Schr\"odinger picture  are defined as time-independent functions and operators, respectively. In 
  our earlier work \cite{Frick1}, we generalized this transition to the transformation 
\begin{equation}
\label{1.1}
 S(t,{\bf x})=\exp[-i(tH-{\bf x}\cdot{\bf P})],
\end{equation}   
where ${\bf P}$ is the momentum operator of the particle. In this context, the functions in the 
Heisenberg picture  and the operators in the Schr\"odinger picture are independent  of the time  
and   space coordinates t, ${\bf x}$. The Fourier transform  of the state  in the  Heisenberg picture  
must be independent of the space-time 
coordinates. That is why the  plane waves $\sim\exp(i{\bf x}\cdot{\bf p})$ cannot be 
applied  in this Fourier transformation. Accordingly, the momentum and the Hamiltonian 
operator of the particle cannot be expressed in terms of the spatial derivative $-i\nabla_{\bf x}$.  Under these  premises, 
there is no ${\bf x}$-representation.  As a result, the  plane 
waves in the new  Schr\"odinger picture  and  also the space-time coordinates in the operators of the new
Heisenberg picture  appear in different representations. In the Heisenberg picture at first one can use the momentum representation  and 
subsequently the representation which is defined via a space-time independent Fourier 
transformation. 
   
 Let the function $\Psi^{(s)}({\bf p})$ be a relativistic wave function of a particle in the momentum representation (${\bf p}$ = momentum, m = mass, $p_0:=\sqrt{m^2+{\bf p}^2}$,  s = spin ). In the context of the generalization $S(t){\Rightarrow}S(t,{\bf x}),$ the function ${\Psi}^{(s)}_{\sigma}({\bf p}$) is a wave function in the Heisenberg picture. Under the  Lorentz transformation $g$  with boost and rotation generators \cite{Wigner1,Wigner2,Bargmann,Shirokov}   
\begin{equation}
\label{1.2}
{\bf N}({\bf p},{\bf
 s}):=ip_0{\nabla}_{\bf p}-\frac{{\bf s}\times{\bf p}}
{p_0+m},\quad{\bf J}({\bf p},{\bf s}):=-i{\bf p}\times{\bf \nabla}_{\bf p}
+{\bf s}:={\bf L}({\bf p})+{\bf s},                    
\end{equation}
 and parameters ${\bf u},u_{0}$, with $({\bf u}^2-u^2_{0}=1$), the function  ${\Psi}^{(s)}_{\mu}({\bf p}$)  transforms by the unitary  representation ($\mu$= spin projection )
\begin{equation}
\label{1.3}
T_{g}{\Psi}^{(s)}_{\mu}({\bf p})=\sum_{\mu^{'}=-s}^{s}W_{\mu\mu^{'}}^{(s)}({\bf p},{\bf u})\,\Psi_{\mu^{'}}^{(s)}(g^{-1}{\bf p}),
\end{equation}
where $W_{\mu\mu^{'}}^{(s)}({\bf p},{\bf u})$ are the Wigner functions (${\bf {\sigma}}$ are the Pauli matrices)
\begin{equation}
\label{1.4}
W^{(1/2)}({\bf p},{\bf u})=\frac{(p_0+m)(u_{0}+1)-{\bf u}\cdot{\bf p}+i{\bf \sigma}({\bf p}\times{\bf u})}{\sqrt{2(u_{0}+1)(p_0+m)(p_{0}u_{0}-{\bf p}\cdot{\bf u}+m)}},\quad{W}^{(1/2)}({\bf p},{\bf u})\,
{W}^{\dagger(1/2)}({\bf p},{\bf u})=1.
\end{equation}
 Such a wave function has  positive definite norm
\begin{equation}
\label{1.5}
\int\frac{d{\bf p}}{p_0}\,\sum_{\mu=-s}^{s}|{\Psi}^{(s)}_{\mu}({\bf p})|=\int\frac{d{\bf p}}{p_0}\,\sum_{\mu=-s}^{s}|{\Psi}^{(s)}_{\mu}({\bf p},t,{\bf x})|<\infty,
\end{equation}
  and can be expanded with respect to irreducible unitary representations of the Lorentz group. This function is  covariant only with respect to the set of spin and momentum variables, and not with respect to each of them separately.
 In (\ref{1.5}), the function $\Psi^{(s)}({\bf p},t,{\bf x}):=S(t,{\bf x})\,\Psi^{(s)}({\bf p})$ is the wave function in the new  Schr\"odinger picture in momentum representation. 
 The relativistic spinors,  transforming by non unitary finite representations of the Lorentz group, have not definite norms. In this case these spinors  are not  useful.

 The unitary  representations correspond to the eigenvalues $1+\alpha^2-{\lambda}^2$  of the first $C_1({\bf p}):={\bf N}^2-{\bf J}^2$  and the eigenvalues $\alpha\lambda$ of the second Casimir operator $C_2({\bf p})={\bf N}\cdot{\bf J},\,(0\leq\alpha<\infty,\quad\lambda=-s,...,s)$. The range of $\alpha$ defines the fundamental series of the  unitary representations.  The formalism of harmonic analysis on the Lorentz group  has  been used by many authors ( a detailed list of references can be found  in Ref.\cite{Joos,Kad,Ruhl,Barut} e.g.). The four-dimensional generalization of the  Heisenberg/Schr\"odinger picture  introduces new features into the nature of the description of particle states. It is necessary to develop a mathematical formalism in the framework of this approach for describing the relativistic particles.
 In  \cite{Frick1,Frick2} the space-time independent  expansion with respect to the unitary irreducible 
representation of the Lorentz group  was  applied as Fourier transformation in the 
Heisenberg picture for the relativistic particle with spin 0 and spin 1/2. This procedure will be applied here for such a practically important example as the  massive particle with spin 1. We shall show that  the consistent determination of the wave functions of a particle in the  Schr\"odinger picture requires the use of the wave functions in the  momentum representation or of the matrix elements of the fundamental series of unitary representations of the Lorentz group.
 Since the  operators $-i\nabla_{\bf x}$  are not momentum operators, we want to find the Hamiltonian and the momentum operators for the particle with spin 1 . In this paper, the operators we obtain are expressed through the group parameter ${\alpha}$.  At first  we will give a short review for the description of particles with spin 0   in 
the context of the  application  of (\ref{1.1}).

\section{Spin 0 - particle} 
 
 In this case,  the  operator $C_1({\bf p})$ have the eigenfunctions 
\begin{equation}
\label{2.1}
\xi^{(0)}({\bf p},{\alpha},{\bf n}):=[(p_{0}n_{0}-{\bf p}\cdot{\bf n})/m]^{-1+i\alpha},
\end{equation}
where  $(n_{0},\,{\bf n})$ is a null vector $n_{0}^{2}-{\bf n}^{2}=0$.\\ The Fourier transforms
 for the states of the relativistic particle 
with spin 0 in terms the basis functions  (\ref{2.1}) $({\bf n}=(\sin{\theta}\cos{\varphi},\sin{\theta}\sin{\varphi},\cos{\theta}),$ $\quad{d\omega}_{\bf n}=\sin{\theta}d{\theta}d{\varphi}$,  have the form \cite{Shapiro},
\begin{equation}
\label{2.2}
\Psi^{(0)}({\bf p})=\frac{1}{(2\pi)^{3/2}}\int{\alpha}^2d{\alpha}\,d{\omega}_{\bf n}\,\Psi^{(0)}({\alpha},{\bf n})\,\xi^{(0)}({\bf p},{\alpha},{\bf n}),
\end{equation}
\begin{equation}
\label{2.3}
\Psi^{(0)}({\alpha},{\bf n})=\frac{1}{(2\pi)^{3/2}}
\int\frac{d{\bf p}}{p_0}\,\Psi^{(0)}({\bf p})\,\xi^{\ast(0)}({\bf p},{\alpha},{\bf n}).
\end{equation}
The functions \(\Psi^{(0)}({\bf p})\) and \(\Psi^{(0)}({\alpha},{\bf n})\)
are the state functions of the particle in \({\bf p}\)- and in the (\({\alpha},{\bf n}\))-representation. The completness and orthogonality relations for the functions $\xi^{(0)}({\bf p},\alpha,{\bf n}),\quad\xi^{\ast(0)}({\bf p},\alpha,{\bf n})$ are given in the Appendix. In the (\({\alpha},{\bf n}\))-representation the free Hamiltonian  and the momentum operators for a particle
with spin 0 are the differential-difference operators \cite{Kad} $({\bf L}:={\bf L}({\theta},{\varphi}))$
\begin{equation}
\label{2.4}
H^{(0)}({\alpha},{\bf
  n})=m\left[\cosh(i\frac{\partial}{{\partial}{\alpha}})+\frac{i}{{\alpha}}\sinh(i\frac{\partial}{{\partial}{\alpha}}) +\frac{{\bf L}^2}{2{\alpha}^2}\exp(i\frac{\partial}{{\partial}{\alpha}})\right],
\end{equation} 
\begin{equation}
\label{2.5}
{\bf P}^{(0)}({\alpha},{\bf
  n})={\bf n}\left[H^{(0)}({\alpha},{\bf
  n})-m\exp(i\frac{\partial}{{\partial}{\alpha}})\right]-m\frac{{\bf n}{\times}{\bf L}}{\alpha}\exp(i\frac{\partial}{{\partial}{\alpha}}).
\end{equation}
The eigenfunction of this operators is $\xi^{(0)}_{\bf p}({\alpha},{\bf n}):=\xi^{\ast(0)}({\bf p},{\alpha},{\bf n})$: 
\begin{equation}
\label{2.6}
  H^{(0)}({\alpha},{\bf
  n})\,\xi^{(0)}_{\bf p}({\alpha},{\bf n})=p_{0}\,\xi^{(0)}_{\bf p}({\alpha},{\bf n}),\quad{\bf P}^{(0)}({\alpha},{\bf
  n)}\,{\xi}^{(0)}_{\bf p}({\alpha},{\bf n})={\bf p}\,{\xi}^{(0)}_{\bf p}({\alpha},{\bf n}).    
\end{equation}

The operators in (\ref{2.4})-(\ref{2.6}) are used for the  relastivistic description of the two-body problem \cite{Kad,Ska,Amir,Drenska,Kag}.  In this case  the vector ${\bf q}={\alpha}{\bf n}/m$ is 
applied.  
In the nonrelativistic limit 
\begin{equation}
\label{2.7}
C_1^{(0)}({\bf p})\rightarrow{-m^2{\bf \nabla}^{2}_{\bf p}},\quad{\xi}^{(0)}({\bf p},{\alpha},{\bf n}){\;}\rightarrow{\;}\exp(-i\alpha{\bf n}\cdot{\bf p}/m),
\end{equation}
\begin{equation}
\label{2.8}
 H^{(0)}(q,{\bf n})-m{\quad}\rightarrow{\quad}-\frac{1}{{2m}{q}^2}\frac{\partial}{{\partial}{q}}{q}^{2}\frac{\partial}{{\partial}{q}}+\frac{\bf L^{2}}{2mq^2},{\quad}{\bf P}^{(0)}(q,{\bf n})\rightarrow{\quad}-i\nabla_{\bf q}.
\end{equation}
The functions $\exp(i{\alpha}{\bf n}\cdot{\bf p}/m)$   realize the unitary irreducible representations of the Galileo group:
\begin{equation}
\label{2.9}
\Psi(\alpha{\bf n})={\frac{1}{{(2\pi)}^{3/2}}}\int{d{\bf p}}\Psi({\bf p})\exp(i{\alpha\bf n}\cdot{\bf p}/m).   
\end{equation}
Since Wigner,  particle are associated with unitary representation of the Poincar\'e group.
 If one introduces the generators of the Lorentz algebra ${\bf N}(\alpha,{\bf n})$ for the particle with spin 0

\begin{equation}
\label{2.10}
{\bf N}^{(0)}(\alpha,{\bf n}):=\alpha{\bf n}+({\bf n}\times{\bf L}+{\bf L}\times{\bf n})/2,
\end{equation}  
then \cite{Frick1}, instead of the vector $\alpha{\bf n}$, the $(\alpha,{\bf n})$-representation can be recognized as a representation of the 
Poincar\'e group. The operators $H^{(0)}(\alpha,{\bf
  n}),\,{\bf P}^{(0)}(\alpha,{\bf
  n}),\,{\bf N}^{(0)}(\alpha,{\bf n}),\,{\bf L}$ satisfy the commutation relations of 
the Poincar\'e  algebra.

The Casimir operator $C_1({\bf p})$, and the functions  $\xi^{(0)}({\bf p},\alpha,{\bf n})$ do not depend on the 
space-time coordinates ${\bf x}$, t. That is why the functions  in the  expansions (\ref{2.2}), (\ref{2.3}) and 
the operators in (\ref{2.4})-(\ref{2.6})  are independent on the space-time  coordinates likewise. In 
the framework of the  four-dimensional generalization of the Heisenberg to the 
Schr\"odinger picture $S(t){\Rightarrow}S(t,{\bf x})$,  the functions in (\ref{2.2}), (\ref{2.3}) and the operators in the (\ref{2.4})-(\ref{2.6})  must be  seen  as 
 functions in the Heisenberg picture  and, accordingly, as operators in the Schr\"odinger 
picture  for the particles with spin 0. If the transformation (\ref{1.1})  is not applied, then we have 
no possibility  to introduce the  plane wave $\exp(i{\bf x}\cdot{\bf p})$ into the relativistic state functions (\ref{2.2}), (\ref{2.3}).   In 
the nonrelativistic limit  there is such a possibility.  The function $\exp(i{\alpha\bf n}\cdot{\bf p}/m)$  in (\ref{2.9})  has the 
form of the plane wave $\exp(i{\bf x}\cdot{\bf p})$. Thus, if the generalization (\ref{1.1}) is not applied,  
the function $\exp(i{\alpha\bf n}\cdot{\bf p}/m)$ can be replaced by $\exp(i{\bf x}\cdot{
\bf p})$.  In such a form, the 
plane waves can be introduced in the Schr\"odinger picture through the Fourier transformation, and then an 
{\bf x}-representation can be introduced.  In the relativistic  expansion (\ref{2.3}) this method cannot be used. The 
functions $\exp(i{\bf x}\cdot{\bf p}),\quad\xi^{\ast(0)}({\bf p},{\alpha},{\bf n})$ have different forms.

 The application of the transformation  (\ref{1.1})  gives the 
state of the particle in the Schr\"odinger picture in ${bf p}$  and $(\alpha,{\bf n})$ -representation:\\
 $\Psi^{(0)}({\bf p},t,{\bf x})=S(t,{\bf x})\Psi^{(0)}({\bf p}),\quad\Psi^{(0)}(\alpha,{\bf n},t,{\bf x})=S(t,{\bf x})\Psi^{(0)}(\alpha,{\bf n})$.

The Fourier expansion $(\exp[-i(p{\cdot}x)]:=\exp[-i(tp_{0}-{\bf x}\cdot{\bf p})])$  
\begin{equation}
\label{2.11}
\Psi^{(0)}({\alpha},{\bf n},t,{\bf x})=\frac{1}{(2\pi)^{3/2}}
\int\frac{d{\bf p}}{p_0}\,\Psi^{(0)}({\bf p})\,{\xi}^{(0)}_{\bf p}({\alpha},{\bf n})\,\exp[-i(p\cdot{x})],
\end{equation}                                 
in contrast to the usual expansion 
\begin{equation}
\label{2.12}
 \Psi^{(0)}(t,{\bf x})=\frac{1}{(2\pi)^{3/2}}
\int\frac{d{\bf p}}{p_0}\Psi^{(0)}({\bf p})\,\exp[-i(p\cdot{x})],
\end{equation}                                                             
contains   the matrix elements  ${\xi}^{(0)}_{\bf p}({\alpha},{\bf n})$ of the unitary representation of the Lorentz group.

In the  expansion (\ref{2.12}), the plane waves in the form  ${\sim}$ const$\exp-i(p{\cdot}x)$ appear as the wave functions of the particle with the definite momentum ${\bf p}$ and spin  0.  Similar  expansion for the particle with spin 1/2 or spin 1  contain the Dirac bispinor (spin 1/2)  or the unit  polarization 4- vector (spin 1), respectively.  
An important difference between (\ref{2.11}) and (\ref{2.12}) is that the plane waves ${\sim}$ const$\exp[-i(p{\cdot}x)]$ 
 without the wave functions in the Heisenberg picture in accordance with the transformation (\ref{1.1})   cannot express the wave functions of the particle.

 The  wave functions with  definite momentum  in the Schr\"odinger picture in $(\alpha,{\bf n})$-representation  are  the functions 
\begin{equation}
\label{2.13}
\xi^{(0)}_{\bf p}(\alpha,{\bf n},t,{\bf x})=\,\xi^{(0)}_{\bf p}(\alpha,{\bf n})\exp[-i(p{\cdot}x)],
\end{equation}
in (\ref{2.11}). 

 The  expression  (2.12) and  the similar  expansions  for the particle with spin 1/2 or spin 1  containing the Dirac bispinor   or the unit  polarization 4- vector   are not
 transformations from one representation to another.
   
In the nonrelativistic limit for the Fourier expansion in  the Schr\"odinger picture we have 
\begin{equation}
\label{2.15}
\Psi(\alpha{\bf n},t,{\bf x})=\frac{1}{(2\pi)^{3/2}}\int{d{\bf p}}\Psi({\bf p})\exp(i{\alpha\bf n}\cdot{\bf p}/m)\exp[-i(t{p^2/2m}-{\bf x}\cdot{\bf p})].
\end{equation} 

\section{Spin 1 - particle}

 The expansions (\ref{2.2}), (\ref{2.3}) are generalized in \cite{Chou,Popov}   for the particle with spin.  They can be expressed in the form
\begin{equation}
\label{3.1}
\Psi^{(s)}_{\mu}({\bf p})=\frac{1}{(2\pi)^{3/2}}\sum_{\mu^{'}=-s}^{s}\int({\mu^{'}}^2+{\alpha}^2)d{\alpha}\,d{\omega}_{\bf
  n}\,D^{(s)}_{\mu{\mu}^{'}}(R_{w})\,\xi^{(0)}({\bf p},\alpha,{\bf n})\,
\Psi^{(s)}_{\mu^{'}}(\alpha,{\bf n}),
\end{equation}
\begin{equation}
\label{3.2}
\Psi^{(s)}_{\mu}(\alpha,{\bf n})=\frac{1}{(2\pi)^{3/2}}\sum_{\mu^{'}=-s}^{s}\int\frac{d{\bf p}}{p_0}\,D^{\dagger(s)}_{\mu{\mu}^{'}}(R_{w})\,\xi^{\ast(0)}({\bf p},\alpha,{\bf n})\,\Psi^{(s)}_{\mu^{'}}({\bf p}),
\end{equation}
 where $\Psi^{(s)}_{\mu}(\alpha,{\bf n})$ is the wave function in $({\alpha},{\bf n}$) represantion and the matrix ${D}^{(s)}(R_w)$ must have  the qualities  of the Wigner rotation in (\ref{1.3}), (\ref{1.4}).
 
 In  \cite{Frick2}  this matrix (spin=1/2) has been found by means of the solutions of the eigenvalue equations of the operator $C_1({\bf p})$
\begin{equation}
\label{3.3}
{D}^{(1/2)}(R_w):={D}^{(1/2)}({\bf p},{\bf n})=\frac{p_0-{\bf p}\cdot{\bf n}+m-i{\sigma}\cdot({\bf p}{\times}{\bf n})}{\sqrt{2(p_0+m)(p_0-{\bf p}\cdot{\bf n})}},\quad{D}^{(1/2)}({\bf p},{\bf n})\,
{D}^{\dagger(1/2)}({\bf p},{\bf n})=1.
\end{equation}
  In the $({\alpha},{\bf n})$-representation   the functions:
\begin{equation}
\label{3.4}
\xi^{(1/2)}_{\bf p}(\alpha,{\bf n}):= D^{\dagger(1/2)}({\bf p},{\bf n})\,\xi^{(0)}_{\bf p}(\alpha,{\bf n}),
\end{equation}                            
were determined as the eigenfunctions of the Hamiltonian and the momentum operator for the particle with spin 1/2.
In this case  
\begin{equation}
\label{3.5}
{\bf J}:={\bf L}+{\bf s},\quad{\bf N}:=\alpha{\bf n}+({\bf n}\times{\bf J}+{\bf J}\times{\bf n})/2.
\end{equation}
and $C_1({\alpha},{\bf n})=1+\alpha^2-({\bf s}\cdot{\bf n})^2,\quad{C}_2({\alpha},{\bf n})=\alpha{\bf s}\cdot{\bf n}.$

 For the  particle with spin one,  we use  the eigenfunctions ${\xi}^{(1)}({\bf p},{\alpha},{\bf n})$  of  both Casimir operators $C_1({\bf p})$ and $C_2({\bf p})$:
 
\begin{equation}
\label{3.6}
{\xi}^{(1)}({\bf p},{\alpha},{\bf n})=D^{(1)}({\bf p},{\bf n})\,D({\bf n})\,{\xi}^{(0)}({\bf p},{\alpha},{\bf n}).
\end{equation}
 The matrix $D^{(1)}({\bf p},{\bf n})$  can be obtained  from the matrix  (\ref{3.3}) and the Clebsch-Gordan coefficients. The matrix  $D^{\dagger}({\bf n})$, \,($D({\bf n})D^{\dagger}({\bf n})=1$) contains the eigenfunctions of the operator ${\bf s}\cdot{\bf n}$, with the eigenvalues $\lambda$=-1, 0, 1.

 In order to define the Hamiltonian  and the momentum operators, we consider the  functions
\begin{equation}
\label{3.7}
\xi^{(1)}_{\bf p}({\alpha},{\bf n}):=D^{\dagger}({\bf n})\,D^{\dagger(1)}({\bf p},{\bf n})\,\xi^{(0)}_{\bf p}({\alpha},{\bf n}),
\end{equation}
as states of the free particle  with spin 1 with a definite momentum in the Heisenberg picture in  $({\alpha},{\bf n})$-representation
\begin{equation}
\label{3.8}
 H^{(1)}({\alpha},{\bf
  n})\,{\xi}^{(1)}_{\bf p}({\alpha},{\bf n})=p_0\,{\xi}^{(1)}_{\bf p}({\alpha},{\bf n}).
\end{equation} 
 The operators in (\ref{3.5}) must be transformed according to the rule
$D^{\dagger}({\bf n})\,{\bf J}\,D({\bf n})=\widetilde{\bf J}$.
             
Applying  (\ref{2.4}), (\ref{2.5}), we can express the functions ${\xi}^{(1)}_{\bf p}({\alpha},{\bf n})$  by means  of the operators
\begin{eqnarray}
\label{3.9}
A({\alpha},{\bf n}):&=&\left[1-\frac{i}{\alpha}\tau-\frac{1+i\alpha+\tau}{\alpha(\alpha-i)}2{\bf s}\cdot{\bf L}+\frac{{i}{\tau}+\alpha}{\alpha^2({\alpha}-{i})}{\bf L}^2-\frac{{2}}{\alpha({\alpha}-{i})}({\bf s}\cdot{\bf L})^2\right]\exp(i\frac{\partial}{\partial\alpha})\nonumber\\&&+\left[1+\frac{i}{\alpha}{\tau}\right]\exp(-i\frac{\partial}{\partial\alpha})+2-\frac{2i}{\alpha-i}{\bf s}\cdot{\bf L},
\end{eqnarray}

\begin{equation}
\label{3.10}
\xi^{(1)}_{\bf p}(\alpha,{\bf n})=D^{\dagger}({\bf n})\,A({\alpha},{\bf n})\,\frac{m}{2}\frac{\xi^{(0)}_{\bf p}({\alpha},{\bf n})}{p_{0}+m},
\end{equation}
where $\tau:=1-({\bf s}\cdot{\bf n})^2$.
Using the equation 
\begin{equation}
\label{3.11}
 H^{(1)}({\alpha},{\bf
  n})\,D^{\dagger}({\bf n})\,A({\alpha},{\bf n})=D^{\dagger}({\bf n})\,A({\alpha},{\bf n})\,H^{(0)}({\alpha},{\bf
  n}),
\end{equation}
we have  

\begin{eqnarray}
\label{3.12}
 H^{(1)}(\alpha,{\bf n})&=&m\left[\cosh(i\frac{\partial}{\partial\alpha})+\frac{i\alpha+{\widetilde\tau}}{\alpha(\alpha-i)}\sinh(i\frac{\partial}{\partial\alpha})+\frac{\widetilde\tau}{\alpha^2}\exp(-i\frac{\partial}{\partial\alpha})+\right.\nonumber\\&&\left.\frac{(\alpha^{2}+\widetilde\tau)\widetilde{\bf J}^{2}}{2\alpha^2(\alpha^2+1)}\exp(i\frac{\partial}{\partial\alpha})-\frac{(\widetilde{\bf s}\cdot\widetilde{\bf L}+2)\widetilde\tau}{\alpha^2+1}-\frac{\widetilde\tau(\widetilde{\bf s}\cdot\widetilde{\bf L}+2)}{\alpha^2}\right].
\end{eqnarray}
In the nonrelativistic limit, with the notation $q={\alpha}/m$, we have
\begin{equation}
\label{3.13}
 H^{(1)}(q,{\bf n})-m{\quad}\rightarrow{\quad}-\frac{1}{{2m}{q}^2}\frac{\partial}{{\partial}{q}}{q}^{2}\frac{\partial}{{\partial}{q}}+\frac{\widetilde{\bf J^{2}}+2[\widetilde{\tau}-(\widetilde{\bf s}\cdot\widetilde{\bf J})\widetilde\tau-\widetilde\tau(\widetilde{\bf s}\cdot\widetilde{\bf J})]}{{2mq^2}}.
\end{equation}
One can determine the momentum operator either by means of the commutation relations of the 
Poincar\'e algebra $[\widetilde{\bf N}, H^{(1)}({\alpha},{\bf
  n})]=-i{\bf P}^{(1)}({\alpha},{\bf
  n})$,
 or by the equations  
\begin{equation}
\label{3.14}
{\bf P}^{(1)}({\alpha},{\bf
  n})\,D^{\dagger}({\bf n})\,A({\alpha},{\bf n})=D^{\dagger}({\bf n})\,A({\alpha},{\bf n})\,{\bf P}^{(0)}({\alpha},{\bf
  n}),
\end{equation}
 \begin{eqnarray}
\label{3.15}
{P}^{(1)}_{3}({\alpha},{\bf
  n})&=&n_3 H^{(1)}({\alpha},{\bf
  n})-m\left[{\bigl(}\frac{\alpha(1-\widetilde{\tau})}{(\alpha^2+1)}+\frac{\widetilde{\tau}}{\alpha}{\bigr)}\exp(i\frac{\partial}{{\partial}{\alpha}})N_3+\right.\nonumber\\&&\left.\frac{1-\widetilde{\tau}+s_3L_3}{\alpha^2+1}\exp(i\frac{\partial}{{\partial}{\alpha}})+\frac{(\widetilde{\bf s}\times{\bf n})_3}{\alpha}(1-\frac{\widetilde{\tau}}{\alpha+i})\right].              
 \end{eqnarray}

The operators $H^{(1)}({\alpha},{\bf
  n}),\,{\bf P}^{(1}({\alpha},{\bf
  n})$ can be 
identified as operators of the massive relativistic spin 1  particle  in the Schr\"odinger picture  in $(\alpha,{\bf n})$-representation.
The state functions in the Schr\"odinger picture can be found by means of the Fourier expansion in the Heisenberg picture (\ref{3.2}) and the transformation (\ref{1.1}):
\begin{equation}
\label{3.16}
{\Psi}^{(1)}_{\mu}({\alpha},{\bf n},t,{\bf x})={\frac{1}{{(2\pi)}^{3/2}}}\sum_{\mu^{'}=-1}^{1}\int\frac{d{\bf p}}{p_0}\,{\xi}^{(1)}_{\bf p\mu\mu^{'}}({\alpha},{\bf n})\,\exp[-i(p\cdot{x})]{\Psi}^{(1)}_{\mu^{'}}({\bf p}).
\end{equation}    
In this case the 
Schr\"odinger equation  is valid
 \begin{equation}
\label{3.17}
i\frac{\partial}{\partial{t}}{\Psi}^{(1)}(\alpha,{\bf n},t,{\bf x})=H^{(1)}(\alpha,{\bf n}){\Psi}^{(1)}(\alpha,{\bf n},t,{\bf x}),
\end{equation}
as well the equation in the spatial derivatives
 \begin{equation}
\label{3.18}
-i{\bf \nabla}_{\bf x}{\Psi}^{(1)}(\alpha,{\bf n},t,{\bf x})={\bf P}^{(1)}({\alpha},{\bf
  n}){\Psi}^{(1)}(\alpha,{\bf n},t,{\bf x}).
\end{equation}

\section{ Partial-wave equations}

 To determine the partial-wave  equations in the $(\alpha,{\bf n})$-representation in Heisenberg picture, we first use the spherical spinors  $\Omega_{{\jmath\ell}m}({\bf n}_{ p}),\quad\Omega_{{\jmath\ell}m}({\bf n})$ being the eigenfunctions of the operators ${\bf s}\cdot{\bf L({\bf p})}$ and ${\bf s}\cdot{\bf L}$. They have the same form as in the nonrelativistic formalism. The equation (\ref{3.8}) permits factorization by introducing the spinors $\widetilde\Omega_{{\lambda\jmath\ell}m}({\bf n}):=D^{\dagger}({\bf n})\cdot\Omega_{{\jmath\ell}m}({\bf n})$. If we introduce $D^{\dagger}_{11}({\bf n})=(1+n_3)/2,\quad{D}^{\dagger}_{10}({\bf n})=-n_{-}/\sqrt{2},\quad{D}^{\dagger}_{1-1}({\bf n})=n^{2}_{-}/2(1+n_3)$,{\quad}$(s_3)_{11}=1$,
then 
 \begin{equation}
\label{4.1}
\widetilde{J}^{(1)}_{3}=L_{3}+s_{3},\quad\widetilde{J}^{(1)}_{-}=L_{-}+s_{3}n_{-}/(1+n_3),\quad\widetilde{J}^{(1)}_{+}=L_{+}+s_{3}n_{+}/(1+n_3),
\end{equation}
 with $\widetilde{{\bf s}\cdot{\bf L}}\,\widetilde{\Omega}_{{\lambda\jmath\ell}m}({\bf n})$=${\beta}\,\widetilde{\Omega}_{{\lambda\jmath\ell}m}({\bf n}),\quad{\beta}$=$\jmath({\jmath})+1-\ell({\ell}-1)-2$.
  
 Let us integrate the expression
\begin{equation}
\label{4.2}
\xi^{(1)}_{\bf p}(\alpha,{\bf n}){\Omega}_{{\jmath\ell}m}({\bf n}_{ p})=D^{\dagger(1)}({\bf n})\,A({\alpha},{\bf n})\,\frac{m}{2}\frac{\xi^{(0)}_{\bf p}({\alpha},{\bf n})}{p_{0}+m}{\Omega}_{{\jmath\ell}m}({\bf n}_{ p}),
\end{equation}
over the angular variables of the ${\bf n}_{ p}$ vectors. The matrix  elements obtained in this way can be written in the form
\begin{equation}
\label{4.3}
\widetilde{A({\alpha},{\bf n})}\,\frac{m}{2}\frac{{\cal{P}}^{(0)}_{l}(\cosh\chi,\alpha)}{p_{0}+m}\cdot4{\pi}i^{l}\widetilde{\Omega}_{{\lambda\jmath\ell}m}({\bf n}.
\end{equation}
We define the partial functions for the particles with spin 1 ${\cal{P}}^{(1)}_{{\lambda\jmath\ell}}(\cosh\chi,\alpha)$ as coefficients that stand in front of $4{\pi}i^{l}\widetilde\Omega_{{\lambda\jmath\ell}m}({\bf n})$ in the expression (\ref{4.3}). These can be expressed in terms of the functions  ${\cal{P}}^{(0)}_{l}(\cosh\chi,\alpha)$ (see Appendix),${\quad}b(\chi):=1/2(\cosh\chi+1)$:

1) for $\jmath=\ell+1$,
\begin{eqnarray}
\label{4.4}
{\cal{P}}^{(1)}_{\lambda\jmath\ell}(\cosh\chi,\alpha)&=&b(\chi)\left[\frac{(\alpha-il)(\alpha-il-i)}{\alpha(\alpha-i|\lambda|)}\exp(i\frac{\partial}{{\partial}{\alpha}})+\frac{2(\alpha-il-i)}{\alpha-i}+\right.\nonumber\\&&\left.\frac{\alpha+i-|\lambda|}{\alpha}\exp(-i\frac{\partial}{{\partial}{\alpha}})\right]{\cal{P}}^{(0)}_{l}(\cosh\chi,\alpha);
\end{eqnarray}

2) for $\jmath=\ell-1$,
\begin{eqnarray}
\label{4.5}
{\cal{P}}^{(1)}_{\lambda\jmath\ell}(\cosh\chi,\alpha)&=&b(\chi)\left[\frac{(\alpha+il)(\alpha+il+i)}{\alpha(\alpha-i|\lambda|)}\exp(i\frac{\partial}{{\partial}{\alpha}})+\frac{2(\alpha+il)}{\alpha-i}+\right.\nonumber\\&&\left.\frac{\alpha+i-|\lambda|}{\alpha}\exp(-i\frac{\partial}{{\partial}{\alpha}})\right]{\cal{P}}^{(0)}_{l}(\cosh\chi,\alpha);
\end{eqnarray}

3) for $\jmath=\ell$ and $|\lambda|=1$,
\begin{eqnarray}
\label{4.6}
{\cal{P}}^{(1)}_{\lambda\jmath\ell}(\cosh\chi,\alpha)&=&b(\chi)\left[\frac{\alpha(\alpha+i)-\jmath(\jmath+1)}{\alpha(\alpha-i)}\exp(i\frac{\partial}{{\partial}{\alpha}})+\frac{2\alpha}{\alpha-i}+\right.\nonumber\\&&\left.\exp(-i\frac{\partial}{{\partial}{\alpha}})\right]{\cal{P}}^{(0)}_{l}(\cosh\chi,\alpha);
\end{eqnarray}

4) for $\jmath=\ell$ and $\lambda=0$,\quad$\widetilde{\Omega}_{{\lambda\jmath\ell}m}({\bf n})$=0,\quad${\cal{P}}^{(1)}_{\lambda\jmath\ell}(\cosh\chi,\alpha)$=0.

For the functions ${\cal{P}}^{(1)}_{\lambda\jmath\ell}(\cosh\chi,\alpha)$ we have
\begin{equation}
\label{4.7} 
H^{(1)}({\alpha},\jmath,\lambda)\,{\cal{P}}^{(1)}_{\lambda\jmath\ell}(\cosh\chi,\alpha)=p_0\,{\cal{P}}^{(1)}_{\lambda\jmath\ell}(\cosh\chi,\alpha),
\end{equation}  
where
 \begin{eqnarray}
\label{4.8}
H^{(1)}({\alpha},\jmath,\lambda):&=&m\left[\frac{\alpha}{2(\alpha-i)}+\frac{\widetilde\tau}{2(\alpha-i)}+\frac{\jmath(\jmath+1)}{2({\alpha}^{2}+1)}(1+\frac{\widetilde\tau}{\alpha^2})\right]\exp(i\frac{\partial}{{\partial}{\alpha}})\nonumber\\&&+m\left[\frac{\alpha-2i}{2(\alpha-i)}(1+\frac{\widetilde\tau}{\alpha^2})\right]\exp(-i\frac{\partial}{\partial\alpha})+m\left(
\begin{array}{ccc}
0&-\frac{\beta+1}{{\alpha}^{2}+1}&0\\-\frac{\beta+2}{{2\alpha}^{2}}&0&-\frac{\beta+2}{{2\alpha}^{2}}\\0&-\frac{\beta+1}{{\alpha}^{2}+1}&0
\end{array}
\right).
\end{eqnarray}

\section{CONCLUSION}
The four-dimensional generalization of the  Heisenberg/Schr\"odinger picture introduces new features into the nature of the description of particle states. In the framework of this approach the plane wave ${\sim}$ const$\cdot\exp[-i(p\cdot{x})]$ in their original sense as the stationary states of a particle cannot appear in the mathematical formalism of the   quantum theory. The consistent determination of the wave functions of a particle in the  Schr\"odinger picture requires the use of the wave functions in the  momentum representation or of the matrix elements of the fundamental series of unitary representations of the Lorentz group.

The found Hamiltonian for spin 1 - particle is a differential-difference operator. The system of eigenfunctions is expressed through  the eigenfunction of the particle with spin zero. 

We hope that the formalism that has been developed here will be employed for solving problems in relativistic quantum physics.
\subsubsection*{ACKNOWLEDGMENTS}

  The author would like to thank Prof. F. W. Hehl for very helpful discussions and comments.

\begin{appendix}
\section*{orthogonality and completeness}

The partial expansion for the  function $\xi^{(0)}({\bf p},\alpha,{\bf n})$ has the form $\\(p_0/m:=\cosh{\chi},\quad{\bf n}_{ p}:={\bf p}/|{\bf p}\mid,\quad{\bf n}_{ p}:=(\sin{\theta}_{ p}\cos{\varphi}_{ p},\sin{\theta}_{ p}\sin{\varphi}_{ p},\cos{\theta}_{ p}),\quad|{\bf p}|/m=\sinh{\chi})$,
\begin{equation}
\label{A.1}
\xi^{(0)}({\bf p},\alpha,{\bf n})=\sum_{l=0}^{\infty}(2l+1)i^{l}{\cal{P}}^{(0)}_{l}(\cosh\chi,\alpha)P_{l}(\bf n_{\bf p}\cdot\bf n),
\end{equation}
with the functions
\begin{eqnarray}
\label{A.2}
{\cal{P}}^{(0)}_{l}(\cosh\chi,\alpha)&=&(-i)^{l}\sqrt\frac{\pi}{2\sinh\chi}\,\frac{\Gamma(i\alpha+l+1)}{\Gamma(i\alpha+1)}{\cal{P}}^{-1/2-l}_{-1/2+i\alpha}(\cosh\chi),
\end{eqnarray}
\begin{eqnarray}
\label{A.3}
{\cal{P}}^{(0)}_{l}(\cosh\chi,\alpha)&=&i^{l}\frac{\Gamma(i\alpha+1)}{\Gamma(-i\alpha+l+1)}({\sinh\chi})^l(\frac{d}{d\sinh\chi})^l{\cal{P}}^{(0)}_{(0)}(\cosh\chi,\alpha),
\end{eqnarray}
\begin{eqnarray}
\label{A.4}
{\cal{P}}^{(0)}_{(0)}(\cosh\chi,\alpha)&=&\frac{\sin(\alpha\chi)}{\alpha\sinh\chi}.
\end{eqnarray}
The orthogonality and completeness conditions for the functions $\xi^{(s)}({\bf p},\alpha,{\bf n})$ have the form
\begin{eqnarray}
\label{A.5}
\frac{1}{(2\pi)^{3}}\sum_{\nu=-s}^{s}\int(\nu^2+\alpha^2)d{\alpha}\,d{\omega}_{\bf n}\,\xi^{\dagger(s)}_{\mu\nu}({\bf p},\alpha,{\bf n})\,
\xi^{(s)}_{\nu\sigma}({\bf p_1},\alpha,{\bf n})&=&\nonumber\\{\delta}_{\mu\sigma}\delta^{(3)}({\bf p}-{\bf p_1})\sqrt{1+{\bf p}^{2}/m^2},
\end{eqnarray}
\begin{eqnarray}
\label{A.6}
\frac{1}{(2\pi)^{3}}\sum_{\nu=-s}^{s}\int\frac{d{\bf p}}{p_0}\,\xi^{(s)}_{\mu\nu}({\bf p},\alpha,{\bf n})\,
\xi^{\dagger(s)}_{\nu\sigma}({\bf p_1},\alpha_1,{\bf n_1})&=&\nonumber\\{\delta}_{\mu\sigma}\frac{\alpha^2}{\mu^2+\alpha^2}\,\delta^{(3)}({\bf n}-{\bf n_1})\,\delta(\alpha-\alpha_1).
\end{eqnarray}
\end{appendix}


\begin{thebibliography}{99}

\bibitem{Frick1} R. A. Frick,  Sov. J. Nucl. Phys. {\bf 38}, 481 (1983).
\bibitem{Wigner1} E. P. Wigner, Ann. of Math. {\bf 40}, 149 (1939).
\bibitem{Wigner2}E. P. Wigner, {\it Group Theory and its Application to the Quantum Mechanics of Atomic Spectra} ( Academic Press 1959).
\bibitem{Bargmann}V. Bargmann and E. P. Wigner, Proc. Natl. Acad.  U.S.A, {\bf 34}, 211 (1948).
\bibitem{Shirokov} Iu. M. Shirokov, Dokl. Akad. Nauk. SSSR. {\bf 94}, 857 (1954); {\bf 99}, 737 (1954); Sov. Phys. JETP {\bf 8}, 703 (1959).
\bibitem{lecture} Lecture in Theoretical Physics, VII-A, eds. W. Brittin, A. O. Barut, (Univ. of Colorado Press, Boulder 1965).
\bibitem{Berestetskii}V. B. Berestetskii, E. M. Lifshitz and L. P. Pitaevskii, {\it Relativistic Quantum Theory, Part 1} (Pergamon Press 1979 ).
\bibitem{Naimark}M. A. Naimark,  {\it Linear Representations of the Lorentz Group} (Pergamon Press, London 1964).
\bibitem{Joos}H. Joos, Fortschritte der Physik {\bf 10}, 65 (1962).

 \bibitem{Kad}V. G. Kadyshevsky, R. M. Mir-Kasimov  and N. B. Skachkov, Nuovo Cimento {\bf A} {\bf 55}, 233 (1968); Sov. J. Part. Nucl. {\bf 2}, 69 
(1973).
\bibitem{Ruhl} W. R\"uhl, {\it The Lorentz group and harmonic analysis}, (N. Y. 1970). 
\bibitem{Barut}A. O. Barut and R. Raczka, {\it Theory of Croup Representations and Applications}\\ ( Warszawa, 1977).
\bibitem{Frick2} R. A. Frick (Frik), JETP Lett. {\bf 39},   89 (1984).
\bibitem{Shapiro} I. S. Shapiro, Sov. Phys. Doklady. {\bf 1}, 91 (1956).
\bibitem{Ska}N. B. Skachkov, I. L.Solovtsov ,  Sov. J. Part. Nucl. {\bf 9}, 1 (1978).
\bibitem{Amir} I. V. Amirkhanov, G. V. Grusha, R. M. Mir-Kasimov,  Sov. J. Part. Nucl. {\bf 12}, 3 (1981).
\bibitem{Drenska}
   S. B. Drenska, S. Sch. Mavrodiev,  Sov. 
J. Part. Nucl. {\bf 15}, 1 (1984).
\bibitem{Kag}E. D. Kagramanov, R. M. Mir-Kasimov, Sh. M. Nagiyev, J. Math. Phys. {\bf 31}, 1733 (1990).
\bibitem{Chou} Chou Kuang-chao and L. G. Zastavenko,  Sov. Phys. JETP {\bf 8}, 990 (1959).
\bibitem{Popov} V. S. Popov, Sov. Phys. JETP {\bf 10}, 794 (1960).
 \bibitem{Frick3} R. A. Frick, N. B. Skachkov. Preprint  JINR. E2-91-449, Dubna (1991).
\end{thebibliography}
\end{document}